\documentclass[12pt]{iopart}

\usepackage{graphicx}
\usepackage{dcolumn}
\usepackage{bm}

\begin{document}

\title[Pairing state at an interface of Sr$_2$RuO$_4$]{Pairing state 
at an interface of Sr$_2$RuO$_4$: parity-mixing, restored time-reversal symmetry,  
and topological superconductivity
}

\author{Y. Tada, N. Kawakami, and S. Fujimoto}

\address{Department of Physics, Kyoto University, Kyoto 606-8502,
Japan}

\newcommand{\vecc}[1]{\mbox{\boldmath $#1$}}

\begin{abstract}
We investigate pairing states realized at the (001) interface of
a spin-triplet superconductor Sr$_2$RuO$_4$ on the basis of 
microscopic calculations. 
Because of a Rashba-type spin-orbit interaction induced at the interface, 
strong parity-mixing of Cooper pairs between a spin-singlet state 
and a spin-triplet state occurs in this system.
There are also strong inter-band pair correlations between
the spin-orbit split bands, in spite of the considerably large spin-orbit splitting.
This is due to frustration between
the spin-orbit interaction and pairing interactions.
In this pairing state, time-reversal symmetry is restored,
in contrast to the bulk Sr$_2$RuO$_4$ which is believed to be a chiral $p+ip$
superconductor with broken time-reversal symmetry.
It is demonstrated that, because of these features, 
the pairing state at the interface is
a promising candidate for the recently proposed 
time-reversal invariant topological superconductor.
\end{abstract}

\maketitle
\section{Introduction}
In unconventional superconductors, 
the BCS order parameter possesses internal degrees of freedom, 
which give rise to various rich physics, 
as explored for Helium 3 \cite{helium}, 
 Sr$_2$RuO$_4$ \cite{maeno,rice}, and  
heavy fermion superconductors \cite{vol,sigristueda}.
For Sr$_2$RuO$_4$, a possible realization of a chiral $p+ip$ pairing state
is suggested by several experimental and 
theoretical studies \cite{rice,kishida,luke}.
In this $p$-wave pairing state, time-reversal symmetry is broken by
orbital degrees of freedom. 
Because of this feature, 
the chiral $p+ip$ 
superconductor bears some similarities to the quantum-Hall-effect (QHE) 
state \cite{stoneroy}. 
For instance, in a chiral $p+ip$ superconductor with open boundaries,
a gapless edge mode propagating in only one direction appear at the boundary edges,
which is in analogy with a chiral edge state in the QHE state.
This similarity ultimately stems from the realization of 
a topological state in both of these quantum condensed phases.
A topological state is a novel class of quantum ground state which 
is characterized not by conventional long-range order, but by 
a topologically nontrivial structure of the Hilbert space.
In a topological state, there is a bulk excitation energy gap
which ensures the stability of this state, and, as mentioned above, 
there are also
gapless edge states which play an important role for transport phenomena.
In the case with broken time-reversal symmetry such as 
the QHE state and the chiral $p+ip$ superconductors,
the topological structure is associated with the existence of a nonzero 
topological number, i.e. the Chern number \cite{lee}.
Recently, another class of a topological state was
theoretically proposed for band insulators \cite{kane,kane2,ber}
and experimentally observed \cite{exp1}.  
This topological state, which is called the $Z_2$ topological insulator, 
possesses time-reversal symmetry, in contrast
to the above-mentioned topological state without time-reversal symmetry, and
is characterized by
the existence of two counter-propagating gapless edge modes, which
are associated with the Kramers doublet.
These gapless edge modes give rise to the quantum spin Hall effect,
which has been attracting recently much interest
in connection with possible applications to spintronics.
As there is similarity between the chiral $p+ip$ superconductors 
and the QHE state,
there is parallelism between the $Z_2$ topological insulator and 
noncentrosymmetric 
$p$-wave superconductors \cite{roy1,roy2,qi,tanaka,satofuji,sch}. 
In these few years,
many classes of noncentrosymmetric superconductors (NCSC), 
the crystal structures of which lack inversion 
symmetry, have been discovered \cite{bauer,akazawa,kimura,sugitani,togano,bad}. 
Some experimental and theoretical 
studies suggest that $p$-wave pairing states may be 
realized in certain systems of NCSC such as CePt$_3$Si and 
Li$_2$Pt$_3$B \cite{haya,yuan,zheng}. 
However, unfortunately, these NCSC are not suitable 
for the realization of the $Z_2$ topological phase, because 
their superconducting gaps possess nodes \cite{izawa,yuan}, 
from which gapless quasiparticles in the bulk appear, and
destabilize the topological state.

In this paper, we investigate a possible realization of the $Z_2$ topological 
superconductivity at an interface of Sr$_2$RuO$_4$.
We consider the $(001)$ interface, at which 
the Rashba-type spin-orbit (SO) interaction breaking inversion symmetry
may be induced \cite{rashba}.
We can also consider a setup in which a thin film of
Sr$_2$RuO$_4$ is fabricated on a substrate 
 with a bias potential applied perpendicular to the $(001)$ interface,
which controls the strength of the Rashba SO interaction.
Apart from the exploration of the $Z_2$ topological superconductivity,
such a system is interesting in that it is suitable for the systematic
investigation on the effect of party-mixing of pairing states
raised by broken inversion symmetry \cite{ede,gor}.
For the realization of substantial parity-mixing of Cooper pairs,
the existence of attractive interactions in both spin-singlet 
and spin-triplet channels is crucially important.
Sr$_2$RuO$_4$ is a good candidate for the realization of such a situation,
because, according to microscopic analysis on the mechanism of 
superconductivity
for this system, both the $p$-wave channel and the $d$-wave channel
enjoy substantially strong attractive interactions \cite{nomura,nomura2}.  
It is expected that the addition of the asymmetric SO interaction
to this system raises the strong admixture of spin-singlet pairs
and spin-triplet pairs. 
In general, the structure of the parity-mixed Cooper pairs is
determined by competition between the asymmetric SO interaction 
and pairing interactions in each channel \cite{ede,frigeri,fuji,fuji2}.
When the asymmetric SO interaction is dominant, 
the structure of the $\mbox{\boldmath $d$}$-vector for the spin-triplet 
component
is mainly constrained by the SO interaction to suppress 
pairings between two SO split bands, which are unfavorable when the SO split
is much larger than the superconducting gap.
However, in the case that the pairing interaction
that is not compatible with the symmetry of 
the asymmetric SO interaction is dominant, 
the direction of the $\mbox{\boldmath $d$}$-vector does not minimize 
the energy cost due to the asymmetric SO interaction, 
yielding inter-band pairings between the SO split bands.
For the case of Sr$_2$RuO$_4$, 
the dominant pairing interaction exists in the $p$-wave channel, 
and the $\mbox{\boldmath $d$}$-vector is perpendicular to the $xy$-plane 
because of the bulk SO interaction of 
the $d$-electron orbitals \cite{rice,nomura,ng,yanase}. 
In this paper, we consider the case that the asymmetric SO interaction
at the interface is stronger than the bulk SO interaction, and examine
how the chiral $p+ip$ state in the bulk is affected by 
the asymmetric SO interaction, and what pairing symmetry 
is most stabilized at the interface. 
This is another purpose of the current paper.

Our main results are as follows.
As the asymmetric SO interaction becomes strong, 
the $\mbox{\boldmath $d$}$-vector of the $p$-wave pairing is oriented to
directions parallel to the $xy$-plane.
However, there exists strong frustration between
the asymmetric SO interaction and 
the $p$-wave pairing interaction because of 
an additional anisotropic structure of the pairing interaction. 
Thus,
the direction of the $\mbox{\boldmath $d$}$-vector does not fully optimize
the asymmetric SO interaction, inducing substantial amount of
Cooper pairs between the two SO split bands, in spite of
the SO splitting much larger than the superconducting gap.
As a result, the stable pairing state possesses $p+d$ wave symmetry, 
rather than
$s+p$ wave symmetry or $d+f$ wave symmetry which is expected to be stabilized 
for the Rashba superconductors when the inter-band pairs 
are suppressed \cite{ede,frigeri,fuji}. 
The notable feature of the $p+d$ wave state is that the single-particle energy
has a full gap, and there are no nodal excitations for small
strength of the asymmetric SO interaction.
Furthermore, since the $\mbox{\boldmath $d$}$-vector is parallel to 
the $xy$-plane, {\it time-reversal symmetry is restored}, which makes sharp
contrast with the chiral superconductivity realized in the bulk of Sr$_2$RuO$_4$.
These two features are quite important 
for the realization of the $Z_2$ topological superconductivity mentioned above.
We can show that the pairing state realized at the interface is topologically
equivalent to the combined state of a $p+ip$ state and a $p-ip$ state,
which is time-reversal invariant, and supports the existence of
counter-propagating gapless edge modes, which carry spin currents.

The organization of this paper is as follows.
In the sections 2 and 3, the pairing state at the interface of Sr$_2$RuO$_4$ is
microscopically investigated on the basis of the scenario that
the pairing interaction is caused by electron correlation effects.
We analyse the structure of the parity-mixed pairing gap.
In the section 4, exploiting the results obtained in the section 3, we present
a possible scenario for the realization of the $Z_2$ topological 
superconductivity in this system.
Discussion and summary are given in the last section.

\section{Model and formulation}\label{sec:model}
In this section, we introduce a low-energy effective model for 
superconductivity realized at an interface of Sr$_2$RuO$_4$, and
present a theoretical framework used for the study on pairing states. 

\subsection{Low-energy effective model}

In Sr$_2$RuO$_4$, there exist three quasi 2-dimensional 
orbitals Ru4d$_{xy,xz,yz}$ in RuO$_2$ plane for electrons
which play the most significant roles for low-energy properties.
There are several theoretical 
proposals for the microscopic origin of pairing interaction
in Sr$_2$RuO$_4$ \cite{maeno,rice}.
One of promising scenarios is that an effective pairing interaction in the 
$p$-wave channel is produced 
by higher order processes of electron-electron interaction through
the Kohn-Luttinger-type mechanism \cite{nomura}.
In this scenario, which is first proposed by Nomura and Yamada, 
among the three bands, $\alpha,\beta$ and $\gamma$-band
formed by the $t_{\rm 2g}$ orbitals, 
the $\gamma$-band originating from the d$_{xy}$ orbital is considered to be
most important for the realization of the superconductivity.
In this paper, we employ this scenario, because this appraoch
enables us to calculate the transition temperature which is quantitatively
in good agreement with experimental observations.
Therefore, to discuss the appearance of superconductivity,
it is sufficient to focus only on the $\gamma$-band.
Although there exists the SO interaction between the $t_{\rm 2g}$ orbitals
(we call this SO interaction the bulk SO interaction)
which tends to direct the $\vecc{d}$-vector parallel to the $z$-axis, 
its effective energy scale for the pinning of the $\vecc{d}$-vector is small
and negligible for a discussion on the transition temperature $T_c$.
However, for electrons near the surface or 
an interface parallel to RuO$_2$ plane, 
there exists another kind of spin-orbit interaction called the 
asymmetric SO interaction
which breaks both reflection symmetry in the momentum space 
$\vecc{k}\rightarrow -\vecc{k}$  and
spin rotation symmetry.
This SO interaction originates from the spin-flip hopping processes
between the Ru $t_{\rm 2g}$ orbitals, of which the wave functions are modulated
by a potential gradient
in the vicinity of an interface.
For the superconductivity, it tends to make the direction
of the $\vecc{d}$-vector perpendicular to the $z$-axis as shown 
in section\ref{sec:result1}.
In the following analysis, we assume that the asymmetric SO interaction is 
sufficiently stronger than the bulk SO interaction, and neglect
effects of the bulk SO interaction.

We, then, simply
describe the electrons of the
$\gamma$-band near an interface by the single band Hubbard model,
\begin{eqnarray}
H&=&
\sum_kc^{\dagger}_k[\varepsilon_k
+\alpha \vecc{\mathcal L}_0(\vecc{k})
\cdot \vecc{\sigma}]c_k
+U\sum_in_{i\uparrow}n_{i\downarrow}, 
\label{eq:action}
\end{eqnarray}
where $c_k=(c_{k\uparrow},c_{k\downarrow})^t$ is the annihilation
operator and $n_{i\sigma}=c_{i\sigma}^{\dagger}c_{i\sigma}$.
The asymmetric type SO interaction induced  near the $(001)$-interface
is incorporated into the first term of the Hamiltonian.
The strength of the asymmetric SO interaction is denoted by $\alpha$.
We assume that the Rashba form of the asymmetric SO interaction \cite{rashba}.
For Sr$_2$RuO$_4$,
the dispersion relation $\varepsilon_k$ and
the Rashba type SO interaction are approximated by  
\begin{eqnarray}
\varepsilon_k&=&-2t_1(\cos k_x+\cos k_y)+4t_2\cos k_x\cos k_y-\mu, \\
\vecc{\mathcal L}_0(\vecc{k})
&=&(\sin k_y,-\sin k_x,0).
\end{eqnarray}
The parameters are fixed as $(t_1,t_2)=(1.0,-0.375)$
taking $t_1$ as the energy unit, and the filling is $n=1.32$.
The hopping integral and the electron density are chosen so that 
the Fermi surface of our model is consistent with the experiments.
In the real system, the form of $\vecc{\mathcal{L}}_0(\vecc{k})$ may be
more complicated. However, as will be shown in the section 3,
this simplified model captures important physics raised by broken
inversion symmetry.

\subsection{Perturbation theory for the Kohn-Luttinger 
mechanism of superconductivity}

There are some theoretical proposals for the mechanism of $p$-wave 
superconductivity
realized in Sr$_2$RuO$_4$.
One promising scenario is that the pairing interaction in this system is
caused by the Kohn-Luttinger mechanism; higher order interaction processes
due to the Coulomb interactions $U$ give rise to 
effective pairing interactions in 
interaction channels with nonzero angular momentum \cite{kohn}.
Actually, Nomura and Yamada demonstrated that interaction processes up 
to the third order in $U$ yield a strong pairing interaction in the $p$-wave channel
for the microscopic model of Sr$_2$RuO$_4$.
This scenario successfully explains the origin of the $p$-wave 
superconductivity
realized in this system.
We, here, apply this perturbation theory for the pairing interaction
to the model (\ref{eq:action}).

For this purpose, we introduce noninteracting Green's function,
\begin{eqnarray}
G_{\alpha \beta}^0(k)&=&\sum_{\tau=\pm 1}l_{\tau \alpha \beta}^0(k)
 G_{\tau}^0(k),\\
l_{\tau \alpha \beta}^0(k)&=&\frac{1}{2}
\left(1+\tau \hat{\vecc{\mathcal L}_0}(\vecc{k})
\cdot \vecc{\sigma} \right)_{\alpha \beta},\\
G_{\tau}^0(k)&=&\frac{1}{i\omega_n-\varepsilon_{k\tau}},
\end{eqnarray}
where $\varepsilon_{k\tau}=\varepsilon_k+
\tau \alpha |\vecc{\mathcal L}_0(\vecc{k})|,
\vecc{\hat{\mathcal L}}_0(\vecc{k})=\vecc{\mathcal L}_0(\vecc{k})/
|\vecc{\mathcal L}_0(\vecc{k})|$
and $|\vecc{\mathcal L}_0(\vecc{k})|=
\sqrt{\sum_{i=1}^3[{\mathcal L}_{0i}(\vecc{k})]^2}$.
$\omega_n$ is the fermionic Matsubara frequency.
As seen in the form of $G_{\alpha \beta}^0$, the Fermi surface
splits into two bands 
whose dispersions are $\varepsilon_+$ and
$\varepsilon_-$ with the splitting $\sim \alpha/v_F$ ($v_F$ is the
averaged Fermi velocity of the two bands).
Note that $\vecc{\mathcal L}_0=0$ at the van Hove points $(0,\pm \pi),
(\pm \pi,0)$ and the Fermi surface is changed little around them 
since the Rashba SO interaction is small there.

The effective pairing interaction is expanded up to the third order with
respect to $U$. It is expressed as,
\begin{eqnarray*}
V_{\sigma_1\sigma_2\sigma_3\sigma_4}(k,k^{\prime})&=&\frac{1}{2}
[V^{{\rm RPA}}_{\sigma_1\sigma_2\sigma_3\sigma_4}(k,k^{\prime})
+V^{{\rm Ver}}_{\sigma_1\sigma_2\sigma_3\sigma_4}(k,k^{\prime})],\\
V^{{\rm RPA}}_{\sigma\bar{\sigma}\sigma\bar{\sigma}}(k,k^{\prime})&=&
U+U^2\chi^0(k+k^{\prime})
+U^3[\left(\chi^0(k-k^{\prime})\right)^2
+\left(\chi^0(k+k^{\prime})\right)^2],\\
V^{{\rm Ver}}_{\sigma\bar{\sigma}\sigma\bar{\sigma}}(k,k^{\prime})&=&
2U^3{\rm Re}\sum_{q}G^0_{\sigma \sigma}(-k+q)[\chi^0(q)
G^0_{\sigma \sigma}(-k^{\prime}+q)
-\phi^0(q)G^0_{\sigma \sigma}(k^{\prime}+q)],\\
V^{{\rm RPA}}_{\sigma\sigma\sigma\sigma}(k,k^{\prime})&=&
-U^2[\chi^0(k-k^{\prime})
-\chi^0(k+k^{\prime})],\\
V^{{\rm Ver}}_{\sigma\sigma\sigma\sigma}(k,k^{\prime})&=&
2U^3{\rm Re}\sum_{q}G^0_{\sigma \sigma}(k+q)
[\chi^0(q)+\phi^0(q)]
[G^0_{\sigma \sigma}(k^{\prime}+q)-G^0_{\sigma \sigma}(-k^{\prime}+q)],\\
V_{\sigma\bar{\sigma}\bar{\sigma}\sigma}(k,k^{\prime})
&=&-V_{\sigma\bar{\sigma}\sigma \bar{\sigma}}(k,-k^{\prime}),\\
{\rm others}&=&0,
\end{eqnarray*}
where
\begin{eqnarray}
\chi^0(q)&=&-\frac{T}{N}
\sum_kG^0_{\sigma \sigma}(q+k)G^0_{\sigma \sigma}(k),\\
\phi^0(q)&=&
-\frac{T}{N}\sum_kG^0_{\sigma \sigma}(q-k)G^0_{\sigma \sigma}(k),
\end{eqnarray}
and $k=(\omega_n,\vecc{k})$,
and $T$ and $N$ are, respectively, temperature and the number of Ru sites.
$\chi^0$ and $\phi^0$ do not depend on spins because 
$G^0_{\uparrow \uparrow}(k)=G^0_{\downarrow \downarrow}(k)$ is
satisfied.
We, here, neglect many terms in $V$
which arise from non-zero off-diagonal elements of
Green's function $G_{\sigma \bar{\sigma}}(k)$, because
$G_{\sigma \bar{\sigma}}(k)$ is smaller than $G_{\sigma \sigma}$
by the factor of $\alpha/\varepsilon_F$ where
$\varepsilon_F$ is the Fermi energy.
Besides, the terms in $V$ with spin-flip processes
represent the perturbative effects of the Rashba SO interaction and, in itself,
do not have crucial importance as long as $\alpha \ll \varepsilon_F$.
Each $V_{\sigma_1\sigma_2\sigma_3\sigma_4}$ consists of 
the RPA-like terms $V^{\rm RPA}$ and the vertex-correction terms $V^{\rm Ver}$. 
The former is included within random phase approximation(RPA), and
the latter is not.
For spin-singlet pairing, the RPA-like terms give dominant attractive 
interaction and play significant roles for its stability,
while the vertex-correction terms do for triplet pairing.

The transition temperatures for the superconductivity are
calculated by solving the linearized Eliashberg equation
\begin{eqnarray}
\lambda \Delta_{\alpha \alpha^{\prime}}(k)&=&-\frac{T}{N}\sum_{k}
V_{\alpha \alpha^{\prime}\beta \beta^{\prime}}(k,k^{\prime})
G^0_{\beta \gamma}(k^{\prime})G^0_{\beta^{\prime}\gamma^{\prime}}(-k^{\prime})
\Delta_{\gamma \gamma^{\prime}}(k^{\prime})
\label{eq:Eliashberg}
\end{eqnarray}
where $\Delta$ is the anomalous self-energy and $\lambda$ is the eigenvalue.
We identify the temperature for which $\lambda (T)=1$ as the transition
temperature.
The normal self-energy is neglected because it is not important 
in the present study.
In this equation, spin-flip processes are included
only in the factor $G(k)G(-k)$ and are not in $V$ within our
approximation.
This is
because the factor $G(k)G(-k)$ behaves like a window function which
allows electrons only near the Fermi surface to participate in 
the superconductivity and therefore has non-perturbative effects of the
Rashba SO interaction,
while the spin-flip scattering processes in $V$ are perturbative.
We note that some of the elements of 
$G^0_{\beta \gamma}(k)G^0_{\beta^{\prime}\gamma^{\prime}}(-k)$
are strongly anisotropic in the $\vecc{k}$-space, 
which restrict the possible symmetries of
the gap functions.
Furthermore, these elements with $\beta\neq\beta'$ or $\gamma\neq\gamma'$
give rise to parity-mixing between spin-singlet and spin-triplet pairs,
which is one of the most remarkable features of NCSC.

The anomalous self-energy is generally written as,
\begin{eqnarray}
\Delta (k)=[D_0(k)
+\vecc{D}(k)\cdot \vecc{\sigma}]i\sigma_2,
\label{eq:D}
\end{eqnarray}
where $D_0(k)$ and $\vecc{D}(k)$ are
the singlet and triplet parts, respectively, and the 
$\vecc{k}$-dependence of $D_{\mu}(i\omega_n,\vecc{k})$ represents 
the symmetry of the superconductivity.
The structure of the $\vecc{d}$-vector 
$\vecc{D}$ is determined by the two factors.
One is the pairing interaction $V$ and, in the case of Rashba 
superconductors, there exists the other factor $\vecc{\mathcal L}_0$.
The microscopic origins of these two factors are generally different, and 
the $\vecc{d}$-vector which $V$ favors does not necessarily coincide with 
the one which $\vecc{\mathcal L}_0$ favors.
For sufficiently large $\alpha$ with $D_{\mu}\ll \alpha \ll \varepsilon_F$,
the most stable direction of the $\vecc{d}$-vector is 
$\vecc{D}\parallel \vecc{{\mathcal L}}_0$ 
because, if
this condition is satisfied, $\Delta(k)$ in the matrix form can be
diagonalized with respect to the $\tau=\pm$ bands.
Conversely, when $\vecc{D}$ is not parallel to $\vecc{{\mathcal L}}_0$,
inter-band pairs between the SO split bands are induced, which generally
lead to pair-breaking effects.
In contrast, for enough small $\alpha$, 
the structure of $\vecc{D}$ is determined by the pairing interaction.
Generally, these two factors which can determine the structure of 
the $\vecc{d}$-vector compete with each other.
In the next section, we discuss the effects of the Rashba SO interaction
on the transition temperature and the structure of $D_{\mu}$.

To solve the Eliashberg equation (\ref{eq:Eliashberg}) numerically,
 we divide the Brillouin zone into 
$64\times 64$ meshes and take 1024 Matsubara frequencies.

\section{Results for the pairing state}\label{sec:result1}

In this section, we present the results for stable pairing states and
their transition temperatures calculated by using the formulation
given in the section 2.

\subsection{Structure of pairing interaction}

\begin{figure}
  \begin{center}
    \begin{tabular}{cc}
      \resizebox{60mm}{!}{\includegraphics{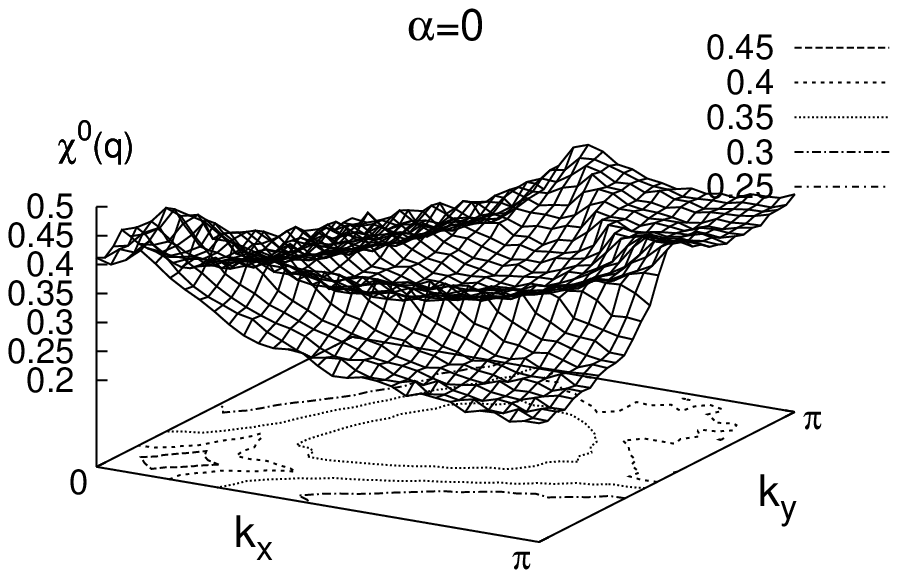}} &
      \resizebox{60mm}{!}{\includegraphics{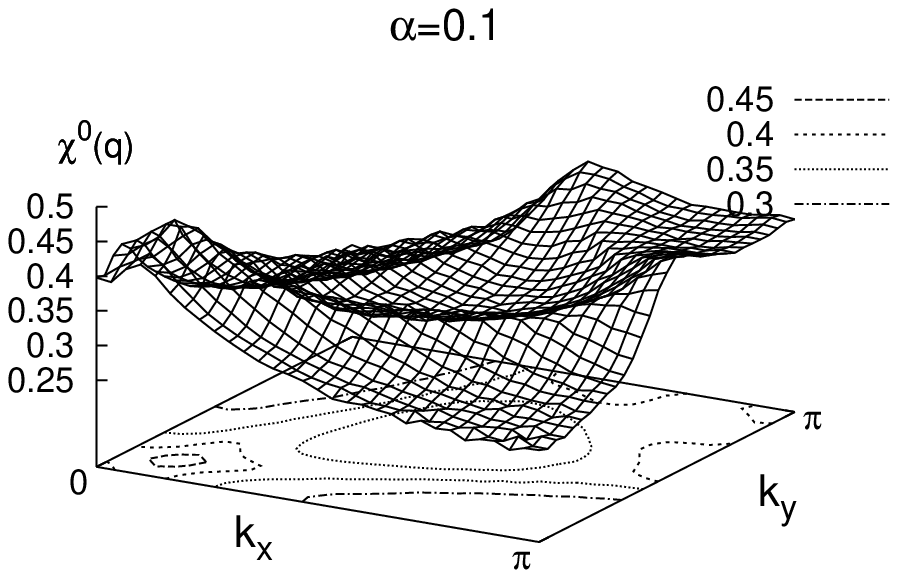}} \\
    \end{tabular}
    \caption{$\chi^0(i\nu_n=0,\vecc{q})$ for $\alpha=0$ (left panel)
    and $\alpha=0.1$ (right panel) at $T=0.01$.}
    \label{fig:chi}
  \end{center}
\end{figure}

\begin{figure}
\begin{center}
\includegraphics[width=4.0in]{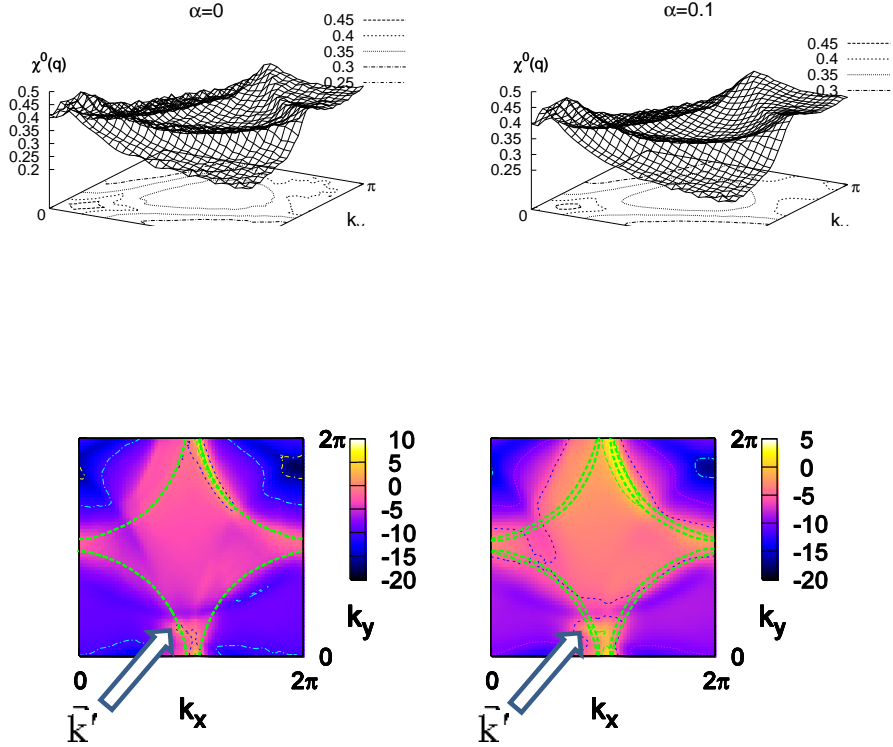}
\end{center}
\caption{\label{fig:Vt}Plot of the pairing interaction $V_t(k,k^{\prime})$ 
on the $\vecc{k}=(k_x,k_y)$ plane with
fixed $\vecc{k}^{\prime}$ 
at $T=0.01,U=5.5$, for $\alpha=0$ (left panel) and $\alpha=0.1$ (right panel).
Here the Matsubara frequencies which 
appear in $V_t$ is fixed as $\nu_n=\nu_n'=i\pi T$. 
$\vecc{k}^{\prime}$ is fixed as illustrated in the figure.}
\end{figure}

We first show the momentum profile of $\chi^0(q)$
in figure \ref{fig:chi} for $\alpha=0$ and $\alpha =0.1$ at $T=0.01$,
which plays an important role for the pairing mechanism. 
The difference in
$\chi^0$ for 
$\alpha=0$ and $\alpha=0.1$ are so small that 
the $\vecc{k}$-dependence of $V$ is also changed little by the Rashba 
SO interaction.
Indeed, we have confirmed that the $\vecc{k}$-dependence of $V$ 
is almost unchanged at least up to $\alpha \sim 0.1$.
This means that, in view of the pairing interaction,  
the most stable symmetry of the superconductivity
for $\alpha=0$ is stable also for $\alpha \neq 0$.
As shown in figure \ref{fig:Vt}, however, 
the amplitude of the pairing interaction for the triplet part
$V_t=\frac{1}{2}
[V^{\rm RPA}_{\sigma \sigma \sigma \sigma}
+V^{\rm Ver}_{\sigma \sigma \sigma \sigma}]$ is decreased for large 
$\alpha$, because the main part of $V_t$ is $V^{\rm Ver}$ which
is sensitive to the electronic structure.
On the other hand, 
the interaction for the singlet part 
$V_s= [V^{\rm RPA}_{\sigma \bar{\sigma} \sigma \bar{\sigma}}
+V^{\rm Ver}_{\sigma \bar{\sigma} \sigma \bar{\sigma}}]$
is not so changed, since $V_s$ is mainly determined by
$V^{\rm RPA}$ which is directly related to 
the $\alpha$-insensitive function $\chi^0$.
Therefore, it is expected that the triplet superconductivity would be
suppressed while the singlet superconductivity unaffected 
through the change in $V$ by
the Rashba SO interaction.
However, as mentioned before, 
the Rahsba SO interaction has the other important effects on $T_c$
which are non-perturbative in the sense that $G(k)G(-k)$
strongly 
restricts the possible symmetries of the gap functions, and
gives rise to parity-mixing of Cooper pairs.

\subsection{Pairing state and transition temperature 
in the case without parity-mixing}

As mentioned above, there are two important effects of the Rashba 
SO interaction
on pairing states: one is to constrain the direction of
the $\vecc{d}$-vector of spin-triplet pairings, and the other is
parity-mixing.
We, first, examine the former effect, neglecting the effect of parity-mixing 
for a while.
Actually, the parity-mixing is not negligible in the present study, 
and will be discussed in the next subsection.

\begin{figure}
\begin{center}
\includegraphics[width=3in]{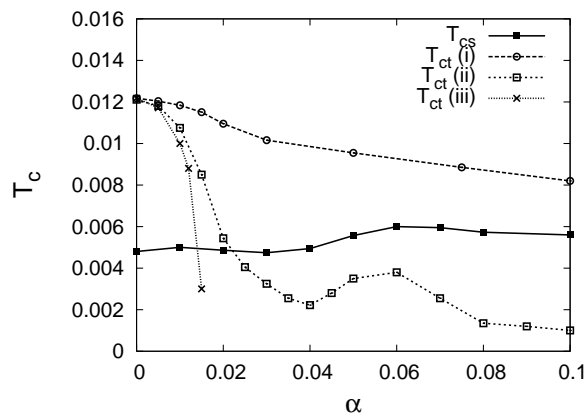}
\end{center}
\caption{\label{fig:Tc_1}$\alpha$ versus $T_{c{\rm s}}$
and $T_{c{\rm t}}$ at $U=5.5$.
The gap functions roughly expressed as $D_0\sim (\cos k_x-\cos k_y)$
for $T_{c{\rm s}}$, 
$\vecc{D}\sim (-\cos k_x\sin k_y \hat{x}+\cos k_y\sin k_x\hat{y})$
for $T_{c{\rm t}}$ (i),
$\vecc{D}\sim [-(a\cos k_x-b\cos k_y)\sin k_y\hat{x}
+(b\cos k_x-a\cos k_y)\sin k_x \hat{y}]$
for $T_{c{\rm t}}$ (ii), and
$\vecc{D}\sim (\cos k_y\sin k_x+i\cos k_x\sin k_y)\hat{z}$
for $T_{c{\rm t}}$ (iii).}
\end{figure}

Neglecting the terms which mix the singlet and the triplet gap functions
in the Eliashberg equation (\ref{eq:Eliashberg}), we calculated
the transition temperatures
$T_c$ for the spin-singlet channels and for the 
spin-triplet channels separately.
We also computed the $\vecc{k}$-dependence of the gap functions self-consistently
from (\ref{eq:Eliashberg}).
Figure \ref{fig:Tc_1} shows the $\alpha$-dependence of $T_c$ 
for the singlet ($T_{c{\rm s}}$) and
the triplet ($T_{c{\rm t}}$) superconductivity at $U=5.5$.
The solid line with closed squares is $T_{c{\rm s}}$ and
the gap function for the spin-singlet pairing 
is roughly given by that with $d_{x^2-y^2}$ symmetry,
$D_0\sim (\cos k_x-\cos k_y)$. 
For the singlet pairing, this is the only one stable gap function,
as in the case without the Rashba SO interaction \cite{nomura}.
The other lines in figure \ref{fig:Tc_1} are for the spin-triplet states, and 
calculated with the assumption that 
the $\vecc{d}$-vector belongs to (i) the A$_1$ representation of the
point group C$_{4v}$, (ii) B$_{1}$ and (iii) E, respectively.
For these representations, the $\vecc{d}$-vector is roughly of the form of 
\begin{enumerate}
\item[(i)]
$\vecc{D}^{{\rm A}_1}
\sim (-\cos k_x\sin k_y \hat{x}+\cos k_y\sin k_x\hat{y})$,

\item[(ii)] $\vecc{D}^{{\rm B}_1}
\sim [-(a\cos k_x-b\cos k_y)\sin k_y\hat{x}
+(b\cos k_x-a\cos k_y)\sin k_x \hat{y}]$ with $a>b$, 

\item[(iii)] $\vecc{D}^{{\rm E}}
\sim (\cos k_y\sin k_x+i\cos k_x\sin k_y)\hat{z}$.

\end{enumerate}
All of these gap functions are $p$-wave gap functions.
Note that they are all different 
from the form $(\cos k_x-\cos k_y)\hat{\vecc{\mathcal L}}_0$
for which the gap function $\Delta$ can be 
diagonalized with respect to the SO split bands
and no inter-band pairing is realized \cite{ede,frigeri,fuji,fuji2}.
Thus, in the spin-triplet pairing states obtained in this calculation,
there are always inter-band Cooper pairs.
This is due to incompatibility between the symmetry of the pairing interactions and
the symmetry of the Rashba SO interaction, as mentioned before.
It should be also notified that the triplet states with $\vecc{D}^{\rm A_1}$ 
and $\vecc{D}^{\rm B_1}$
are time-reversal invariant, 
while the triplet state with $\vecc{D}^{\rm E}$ is not, but a chiral $p+ip$ state.
This is easily seen as follows.
For the state with $\vecc{D}^{\rm A_1}$ or $\vecc{D}^{\rm B_1}$, 
under time-reversal 
operation, the gap function for $\uparrow\uparrow$ pairs 
$\Delta_{\uparrow\uparrow}(\vecc{k})=-D_1(\vecc{k})+iD_2(\vecc{k})$
is transformed as $-D_1(-\vecc{k})-iD_2(-\vecc{k})=
\Delta_{\downarrow\downarrow}(\vecc{k})$, and
also, $\Delta_{\downarrow\downarrow}(\vecc{k})\rightarrow
\Delta_{\uparrow\uparrow}(\vecc{k})$. 
Thus, the $\vecc{D}^{\rm A_1}$ state and
the $\vecc{D}^{\rm B_1}$ state
are time-reversal invariant.

As can be seen in figure \ref{fig:Tc_1}, $T_{c{\rm s}}$ is 
not so strongly affected by 
the Rashba SO interaction. 
In contrast, $T_{c{\rm t}}$ is rapidly decreased, as $\alpha$ increases,
especially for $\vecc{D}^{{\rm E}}$.
The direction of this $\vecc{d}$-vector is not compatible with the
Rashba SO interaction at all.
For this case, the factor $G(k)G(-k)$ in eq.(\ref{eq:Eliashberg})
can be negative 
on a wide area of the Fermi surface.
Therefore, the direction of the $\vecc{d}$-vector strongly tends to align
in the $xy$-plane.
On the other hand,
the superconductivity described by the gap function $\vecc{D}^{{\rm A}_1}$
is most stable for large $\alpha$, because $\vecc{\mathcal L}_0$ in the Rashba SO
interaction,
which tends to direct the $\vecc{d}$-vector so that the condition $\vecc{D}\parallel 
\vecc{\mathcal L}_0$ is satisfied, 
belongs also to A$_1$ irreducible representation.
In this sense, $\vecc{D}^{{\rm A}_1}$ is, to some extent, compatible with
the Rashba SO interaction, though it does not yet fully optimize the
Rashba interaction, leading to strong inter-band pair correlations.
In the case of $\vecc{D}^{{\rm B}_1}\sim [(a\cos k_x-b\cos k_y)\sin k_y\hat{x}
-(b\cos k_x-a\cos k_y)\sin k_x \hat{y}]$, the parameters $a$ and $b$ 
in the gap function changes 
as $\alpha$ is increased.
For $\alpha=0$, $(a,b)\propto (1,0)$. 
When $\alpha$ is turned on,
$a$ is decreased while $b$ increased 
so that $\vecc{D}^{{\rm B}_1}$ 
would become close to the form compatible with $\vecc{\mathcal L}_0$.
This change is continuous with respect to $\alpha$ and 
the gap function $\vecc{D}^{{\rm B}_1}$ is transformed gradually.
As will be shown later in figure \ref{fig:Delta05}
and figure \ref{fig:Delta10},
$\vecc{D}^{{\rm B}_1}$ has $p$-wave like character for small $\alpha$ 
and is gradually changed into the $f$-wave like gap function (i.e. $a=b$) as
$\alpha$ is increased.
As seen in figure \ref{fig:Tc_1}, $T_c$ for $\vecc{D}^{{\rm B}_1}$ has
a minimum around  $\alpha \simeq 0.04$ and a hump around $\alpha\simeq 0.06$.
This $\alpha$-dependence is understood as follows.
As mentioned before, $T_c$ is determined by competition and interplay between
 the pairing interaction and the Rashba SO interaction.
As $\alpha$ increases, because of the change of the electronic structure
due to the Rashba SO interaction, the pairing interaction in the $p$-wave channel
becomes weak in our model. This results in the overall decrease of $T_c$ for
$\vecc{D}^{{\rm B}_1}$ state.
On the other hand, the increase of $\alpha$ also changes the structure of 
the $\vecc{D}^{{\rm B}_1}$ gap function more
compatible with the Rashba SO interaction, suppressing inter-band pairings between
the SO split bands and also associated pair-breaking effects.
A slight increase for $0.04<\alpha<0.06$ is caused by this suppression of
the inter-band pairings.
In our model, for $\alpha>0.1$, the decrease of $T_c$ for $\vecc{D}^{{\rm B}_1}$
is substantial. Thus, the pairing state with $a=b$, i.e. $f$-wave state,
can not be realized.

\subsection{Pairing state and transition temperature in the case
with parity-mixing}

\begin{figure}
\begin{center}
\includegraphics[width=3in]{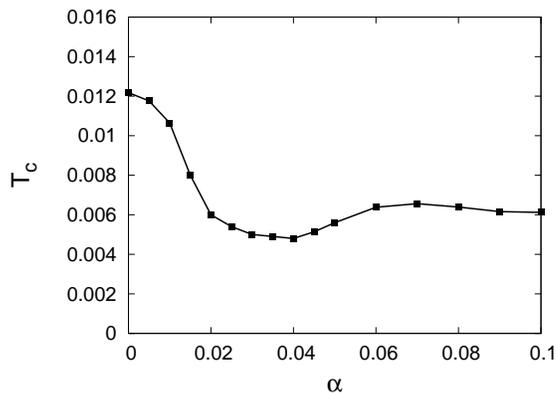}
\end{center}
\caption{\label{fig:Tc_2}$\alpha$ versus $T_{c}$ for the singlet 
and triplet mixed B$_1$ symmetric state at $U=5.5$.}
\end{figure}

The pairing states obtained in the previous subsection is 
drastically changed once we take into account 
the parity mixing of the singlet and the triplet gap functions.
According to the behaviors of $T_{c{\rm s}}$ and $T_{c{\rm t}}$
in figure \ref{fig:Tc_1}, 
the $d$-wave and the $p$-wave
pairing states may be mixed through the Rashba SO interaction.
For the admixture of the gap functions, however, only the gap functions
which belong to the same irreducible representation of the point
group are allowed to coexist.
The $p$-wave gap function 
$\vecc{D}^{{\rm A}_1}$ with the highest $T_{c{\rm t}}$
belongs to A$_1$ representation of C$_{4v}$ and the 
$d$-wave gap function $D_0\sim (\cos k_x-\cos k_y$) to B$_1$.
This implies that these two gap functions can not be mixed.
Then, the next candidate is the admixture of
the $d$-wave state and the $p$-wave state with $\vecc{D}^{{\rm B}_1}$.
The symmetry argument allows this admixture.
Indeed, we found that the only one solution of eq.(\ref{eq:Eliashberg}) 
is the $d+p ({\rm B_1})$ wave state.
Figure \ref{fig:Tc_2} shows the transition temperature $T_c$ for this 
$d+p ({\rm B_1})$ wave superconductivity as a function of $\alpha$.
For small $\alpha$, the superconductivity is dominated by 
the spin-triplet state, while
for large $\alpha$ it is dominated by the spin-singlet state.
For intermediate values of $\alpha$, 
the two gap functions are strongly mixed with the same order of
magnitude.

\begin{figure}
  \begin{center}
    \begin{tabular}{cc}
      \resizebox{62mm}{!}{\includegraphics{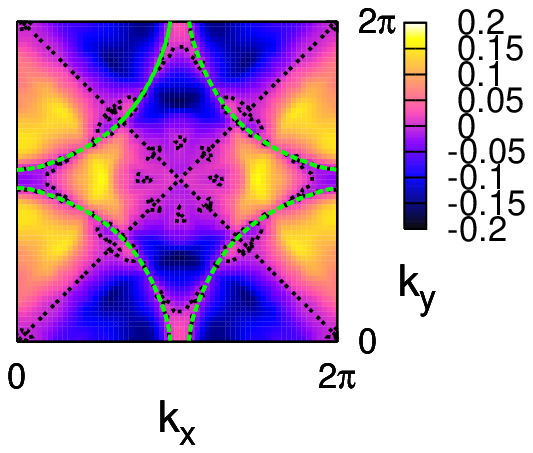}} &
      \resizebox{58mm}{!}{\includegraphics{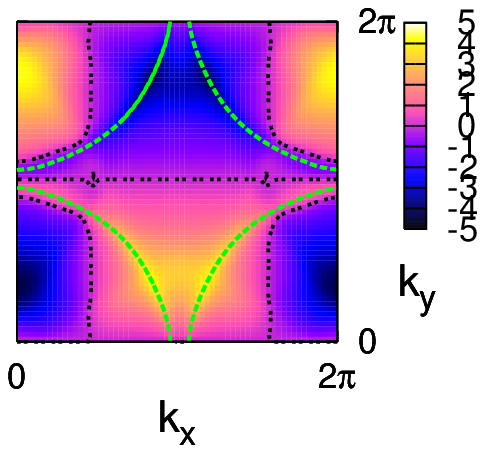}} \\
    \end{tabular}
    \caption{The gap functions for
the $d+p ({\rm B_1})$ state plotted on $(k_x,k_y)$ plane.
$D_0(i\omega_n=i\pi T_c,\vecc{k})$ (left panel)
    and $D_1(i\omega_n=i\pi T_c,\vecc{k})$ (right panel)
    for $\alpha=0.005,U=5.5$. 
Green dotted lines indicate locations of the SO split Fermi surfaces.
Black dotted lines indicate locations of
gap nodes where the gap amplitude vanishes.}
    \label{fig:Delta05}
  \end{center}
\end{figure}

\begin{figure}
  \begin{center}
    \begin{tabular}{cc}
      \resizebox{58mm}{!}{\includegraphics{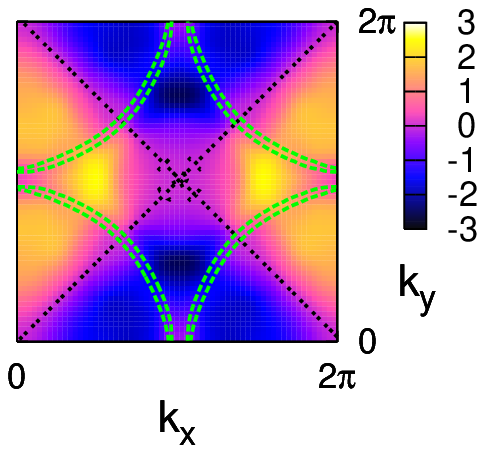}} &
      \resizebox{62mm}{!}{\includegraphics{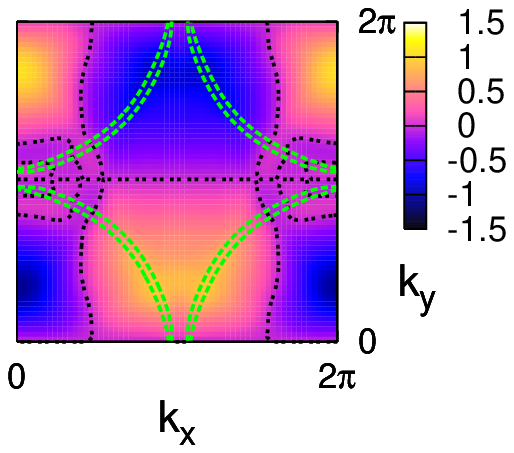}} \\
    \end{tabular}
    \caption{The gap functions for
the $d+p ({\rm B_1})$ state plotted on $(k_x,k_y)$ plane.
$D_0(i\omega_n=i\pi T_c,\vecc{k})$ (left panel)
    and $D_1(i\omega_n=i\pi T_c,\vecc{k})$ (right panel)
    for $\alpha=0.1,U=5.5$. 
Green dotted lines indicate locations of the SO split Fermi surfaces.
Black dotted lines indicate locations of
gap nodes where the gap amplitude vanishes.}
    \label{fig:Delta10}
  \end{center}
\end{figure}

We show the $\vecc{k}$ dependence of $D_0(i\pi T_c,\vecc{k})$ and 
$D_1(i\pi T_c,\vecc{k})$ for the $d+p({\rm B_1})$ state
in figure \ref{fig:Delta05} for $\alpha=0.005$ and 
in figure \ref{fig:Delta10} for $\alpha=0.1$.
Note that $D_2(i\omega_n,k_x,k_y)=D_1(i\omega_n,k_y,k_x)$ is
satisfied for the B$_1$ symmetric superconducting state.
For small $\alpha$, $D_1$ exhibits a conventional $p$-wave behavior, 
in the sense that the Fermi surfaces cross the nodal lines of $D_1$
only near $(\pm \pi,0)$ and the amplitude of $D_1$ 
is much larger than that of $D_0$.
Meanwhile, for large $\alpha$, $D_1$ is
more like a $f$-wave state in the sense that the Fermi surfaces cross 
the nodal lines of $D_1$ 
around $(\pm 0.4\pi,\pm 0.7\pi)$ in addition to $(\pm \pi,0)$,
though the single-particle energy is fully gapped with no nodal lines 
as will be clarified in the next section.
The change from the conventional $p$-wave like gap function to the $f$-wave like
one is continuous, and actually the $f$-wave like state should be classified
as a $p$-wave state with higher harmonics.
In our model, even for large $\alpha>0.1$, 
$\vecc{D}^{\rm B_1}$ does not change to the form $D_0\hat{\vecc{\mathcal L}}_0$
(i.e. a genuine $f$-wave state)
for which no inter-band pairing exists.
This is because $V_t$ does not favor such a structure of the $d$-vector.
The two factors for the determination of the $d$-vector,
the pairing interaction and the Rashba SO interaction, generally have
different origins and favor different types of gap functions.
Therefore, our results imply that,
for the case that the pairing interaction in a spin-triplet channel
is sufficiently strong,
it is rather generally hard for the gap function to be compatible with the 
Rashba SO interaction, resulting in the existence of the inter-band
pairing, in contrast to previous studies on simple models 
in which it is assumed that the spin-triplet 
pairing interaction compatible with the asymmetric
SO interaction always exists \cite{ede,frigeri,fuji}.
In the case that 
the pairing interaction for the triplet superconductivity
is very small compared with that for the singlet one 
and the asymmetric SO interaction, 
the triplet component is induced by the singlet component,
hence $\vecc{D}\parallel \hat{\vecc{\mathcal L}}_0$ can be satisfied.

We note that the relative phase of $\vecc{D}$ to $D_0$
is determined through the Rashba SO interaction.
The eigenvalue of eq.(\ref{eq:Eliashberg}) for $\Delta=
(D_0-\vecc{D}\cdot\vecc{\sigma})i\sigma_2$ where $(D_0,\vecc{D})$
is the gap function illustrated in figures \ref{fig:Delta05} and
\ref{fig:Delta10} is smaller than that for $\Delta=
(D_0+\vecc{D}\cdot\vecc{\sigma})i\sigma_2$.
The resulting gap function has no degrees of freedom with respect to
the relative phase between the singlet component $D_0$ and 
the triplet component $\vecc{D}$.

As mentioned in the previous subsection, the $d+p({\rm B_1})$ pairing state
is time-reversal invariant.
Thus, the asymmetric SO interaction restores the time-reversal symmetry
of the superconducting state,
which is broken in the bulk of Sr$_2$RuO$_4$ where the chiral $p+ip$ state
is realized.
This restored time-reversal symmetry bears an important
implication for the realization of time-reversal invariant topological
superconductivity, as will be discussed in the next section.

\section{A possible realization of topological superconductivity}
\begin{figure}
  \begin{center}
    \begin{tabular}{cc}
      \resizebox{58mm}{!}{\includegraphics{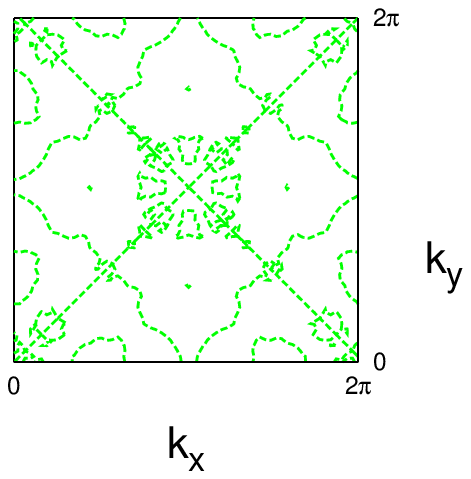}} &
      \resizebox{58mm}{!}{\includegraphics{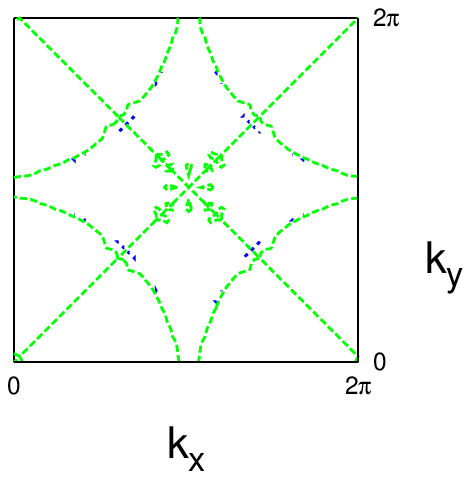}} \\
    \end{tabular}
    \caption{\mbox{\boldmath $k$}-points 
satisfying eqs.(\ref{cond1}) (blue line)
and (\ref{cond2}) (green line) plotted for
$\alpha=0.005$ (left panel) and $\alpha=0.02$ (right panel).
For $\alpha=0.005$, the left-hand side of (\ref{cond1}) is larger than $0$ 
in the entire Brillouin zone.
}
    \label{fig:cond-gap}
  \end{center}
\end{figure}

In this section, we present the scenario of the $Z_2$ topological superconductivity on the basis of the results obtained in the previous section.
As mentioned in the introduction, the stability of
the topological superconductivity is ensured by time-reversal symmetry
and the existence of 
a full energy gap which separates the topologically nontrivial 
ground state from excited states 
\cite{kane,ber,roy1,roy2}; there should be no nodal lines of the gap,
from which gapless excitations may emerge, destabilizing the gapless edge modes
and destroying the topological phase.
The time-reversal invariance is evident for our $d+p({\rm B_1})$ state, 
as mentioned in the previous section.
Thus, we, here, examine whether the single-particle energy for
the $d+p({\rm B_1})$ wave state is fully gapped, and there is no nodal lines of the gap.
To simplify our analysis, we assume the BCS mean field Hamiltonian,
which is, in the matrix form,
\begin{eqnarray}
\left(
\begin{array}{cc}
\varepsilon_k+\alpha\mbox{\boldmath $\mathcal{L}$}_0(\mbox{\boldmath $k$})
\cdot\mbox{\boldmath $\sigma$} & 
\hat{\Delta}_k \\
\hat{\Delta}_k^{\dagger} &
-\varepsilon_k-\alpha\mbox{\boldmath $\mathcal{L}$}_0(-\mbox{\boldmath $k$})
\cdot\mbox{\boldmath $\sigma$}^{t}
\end{array}
\right),
\label{enemat}
\end{eqnarray}
with
\begin{equation}
\hat{\Delta}_k=D_0(\mbox{\boldmath $k$})i\sigma_2
+\mbox{\boldmath $D$}(\mbox{\boldmath $k$})
\cdot\mbox{\boldmath $\sigma$}i\sigma_2,
\end{equation}
where $D_0(\mbox{\boldmath $k$})$ is the spin-singlet gap, and
$\mbox{\boldmath $D$}(\mbox{\boldmath $k$})$ is 
the $\mbox{\boldmath $d$}$-vector for the spin-triplet component.
The energy eigen value of (\ref{enemat}) is obtained as \cite{sato}
\begin{eqnarray}
E_{k\pm}&=&[\varepsilon_k^2+\alpha^2|\mbox{\boldmath $\mathcal{L}$}_0(\mbox{\boldmath $k$})|^2+|\mbox{\boldmath $D$}(\mbox{\boldmath $k$})|^2
+D_0(\mbox{\boldmath $k$})^2 \nonumber \\
&&\pm 2
\sqrt{(\varepsilon_k\alpha\mbox{\boldmath $\mathcal{L}$}_0(\mbox{\boldmath $k$})+D_0(\mbox{\boldmath $k$})\mbox{\boldmath $D$}(\mbox{\boldmath $k$}))^2+(\alpha\mbox{\boldmath $\mathcal{L}$}_0(\mbox{\boldmath $k$})
\times\mbox{\boldmath $D$}(\mbox{\boldmath $k$}))^2}]^{\frac{1}{2}},
\label{eneband}
\end{eqnarray}
and the minus branch of the eigen values $-E_{k\pm}$.
When $\mbox{\boldmath $D$}(\mbox{\boldmath $k$})\parallel
\mbox{\boldmath $\mathcal{L}$}_0(\mbox{\boldmath $k$})$, 
the energy spectrum (\ref{eneband}) is reduced to 
$E_{k\pm}=\sqrt{\varepsilon_{k\pm}^2+\Delta_{\pm}^2(\mbox{\boldmath $k$})}$,
with $\varepsilon_{k\pm}=\varepsilon_k\pm\alpha |\mbox{\boldmath $\mathcal{L}$}_0(\mbox{\boldmath $k$})|$ and
$\Delta_{\pm}(\mbox{\boldmath $k$})=D_0(\mbox{\boldmath $k$})\pm
|\mbox{\boldmath $D$}(\mbox{\boldmath $k$})|$.
The energy spectrum is diagonal with respect to
the SO-split-band-index $\pm$, and  
there are no inter-band Cooper pairs.
For the $p+d$ wave state obtained in the previous section,
$\mbox{\boldmath $D$}(\mbox{\boldmath $k$})\parallel
\mbox{\boldmath $\mathcal{L}$}_0(\mbox{\boldmath $k$})$ does not hold
for a wide range of parameters.   
In this situation, the energy spectrum (\ref{eneband}) is not diagonal
with respect to the band index, which implies that there exist
inter-band Cooper pairs as well as intra-band pairs.
The condition for the existence of gapless excitations,
$E_{k\pm}=0$, is recast in,
\begin{eqnarray}
\varepsilon_k^2-\alpha^2|\mbox{\boldmath $\mathcal{L}$}_0
(\mbox{\boldmath $k$})|^2+|\mbox{\boldmath $D$}(\mbox{\boldmath $k$})|^2-D_0(\mbox{\boldmath $k$})^2=0,
\label{cond1}
\end{eqnarray}
\begin{eqnarray}
\varepsilon_kD_0(\mbox{\boldmath $k$})- 
\alpha\mbox{\boldmath $\mathcal{L}$}_0(\mbox{\boldmath $k$})
\cdot\mbox{\boldmath $D$}(\mbox{\boldmath $k$})=0.
\label{cond2}
\end{eqnarray}
$\mbox{\boldmath $k$}$-points at which the excitation energy
gap closes should satisfy both 
eqs.(\ref{cond1}) and (\ref{cond2}).
We examine these conditions numerically for the $d+p({\rm B_1})$ state.
The calculations presented in the previous sections are valid only for $T \geq T_c$.
Thus, for the evaluation of (\ref{cond1}) which 
requires the magnitude of the gap function, we assume that
the maximum values of the superconducting gaps $D_0$ and 
$|\mbox{\boldmath $D$}(\mbox{\boldmath $k$})|$ at $T=0$
are obtained from the BCS mean field relation 
$\Delta/T_c=1.764$.
Using this approximation, we derive $\mbox{\boldmath $k$}$-points satisfying 
eqs.(\ref{cond1}) and (\ref{cond2}) for the $d+p$(B$_1$) state.
The results for some values of $\alpha$ are shown in
figure \ref{fig:cond-gap}.
We found that when $\alpha$ is sufficiently small,
the left-hand side of (\ref{cond1}) is positive for all $\mbox{\boldmath $k$}$
in the entire Brillouin zone, and thus there are no gapless excitations.
As $\alpha$ is increased from $0$, the left-hand side of (\ref{cond1})
decreases, 
and when $\alpha$ reaches to a value $\alpha_0\sim 0.02$,
eq.(\ref{cond1}) is fulfilled in a certain region of $\mbox{\boldmath $k$}$
where $k_x\approx \pm k_y\approx \pm 0.65\pi$.
The condition (\ref{cond2}) is also satisfied exactly on the line $k_x=\pm k_y$,
because of the B$_1$ symmetry; i.e. $D_1(-k_y,k_x)=D_2(k_x,k_y)$ and $D_0(k_x,k_x)=0$.
This implies that the gap collapses at certain $\mbox{\boldmath $k$}$-points on the line
$k_x=k_y$. 
Our numerical data for the gap function indicate that 
this gap-closing does not occur for $0<\alpha<\alpha_0$.
For this parameter region, the bulk excitations have the full energy gap
in the whole Brillouin zone, which is a necessary condition for the realization of
the topological superconductivity.

We, next, consider the adiabatic deformation of the above $d+p({\rm B_1})$ wave state
to a topologically equivalent state. This deformation is achieved by changing
parameters of the Hamiltonian without closing the bulk gap \cite{kane,kane2}.
Since  any gradual changes of parameters cannot change 
the nonzero topological number which is discrete, 
the existence of the bulk gap ensures the topological stability of the state.
As mentioned above, for $0<\alpha<\alpha_0$,
$\varepsilon_k^2+|\mbox{\boldmath $D$}(\mbox{\boldmath $k$})|^2
>\alpha^2|\mbox{\boldmath $\mathcal{L}$}_0
(\mbox{\boldmath $k$})|^2+D_0(\mbox{\boldmath $k$})^2>0$ holds. 
Thus we can change adiabatically the magnitude of the spin-singlet gap 
$D_0(\mbox{\boldmath $k$})$ and the strength of the SO interaction $\alpha$
to zero without closing the excitation gap.
After this deformation, the system is equivalent to 
a combined system of a $p+ip$ state and a 
$p-ip$ state, which indeed exhibits the $Z_2$ topological superconductivity 
\cite{roy1,qi,tanaka,satofuji}.
As a result, the $d+p({\rm B_1})$ wave state obtained in the section 3 is topologically
equivalent to 
the $Z_2$ topological superconductivity.

In this $Z_2$ topological phase, there are
counter-propagating gapless edge states, 
which are Majorana fermions \cite{roy1,satofuji}.
The Majorana edge states 
may give rise to intriguing transport phenomena associated 
with spin currents \cite{ech}.
Since there is a diamagnetic supercurrent on the boundary surface,
for the detection of the spin current carried by edge quasiparticles, 
thermomagnetic effects may be utilized \cite{satofuji}.
Also, the gapless edge quasiparticles may be observed as a zero bias peak of 
tunneling currents \cite{tanaka,satofuji}.

\section{Discussion and summary}
In this paper, we have investigated pairing states realized at the
$(001)$ interface of Sr$_2$RuO$_4$ by using microscopic calculations based on
the Kohn-Luttinger-type pairing mechanism.
It is found that at the (001) interface of Sr$_2$RuO$_4$, 
the strong admixture of $p$-wave pairings and $d$-wave pairings
realizes, and thus this system is suitable for the exploration of
strong parity-mixing of Cooper pairs caused by
broken inversion symmetry. 
An important implication of our results is as follows.
When there are strong spin-triplet pairing correlations
in NCSC, the frustration between the pairing interaction and the asymmetric SO
interaction occurs quite generally, because of incompatibility between
the pairing interaction and the symmetry of the asymmetric SO interaction.
This yields substantial spin-triplet inter-band pairs 
between electrons in two SO split bands, even when the size of the SO split 
is considerably larger than the superconducting gap.
Because of this feature, the most stable parity-mixed pairing state
realized at the (001) interface of Sr$_2$RuO$_4$ is the $d+p({\rm B_1})$ 
wave state, in which time-reversal symmetry is restored,
in contrast to
the bulk Sr$_2$RuO$_4$, which is believed to be in 
the chiral $p+ip$ state with broken time-reversal
symmetry.
Another intriguing conclusion drawn from our results is that
this $d+p({\rm B_1})$ wave state can be
a promising candidate of
the recently-proposed $Z_2$ topological superconductivity.
That is, the $d+p({\rm B_1})$ wave state is
topologically equivalent to the state that consists of
a $p+ip$ state for $\uparrow\uparrow$ pairs and a $p-ip$ state
for $\downarrow\downarrow$ pairs, which supports
the existence of
counter-propagating gapless edge states carrying spin currents.
This feature may be observed experimentally in the transport properties
for heat currents and spin currents, 
as discussed in some literature \cite{tanaka,satofuji,ech}.

Some concluding remarks are in order.
For the spin-triplet pairing state obtained in the above analysis,
inter-band pair correlation between the SO split bands is substantially
large.
This result for inter-band pairings implies that 
pairing states with center-of-mass momentum
such as the Fulde-Ferrel-Larkin-Ovchinikov (FFLO) state~\cite{fflo,fflo2} 
may be more stabilized for some parameter regions
compared to uniform states considered in the current paper.
The stability of the FFLO state depends on competition between
pairing interaction for this state and the cost of the 
kinetic energy due to the finite center-of-mass momentum.
The issue of a possible realization of the FFLO state at an interface
of Sr$_2$RuO$_4$
is quite intriguing, and should be addressed in the near future.

In the argument for the realization of the topological superconductivity
presented in this paper, we did not consider effects of the bulk SO interaction
due to the $d$-electron orbitals, but simply assumed
that the asymmetric SO interaction overwhelms the bulk SO interaction 
in the vicinity of the interface.
Actually, there should exist a domain boundary between
the chiral $p+ip$ state in the bulk governed by
the bulk SO interaction and the time-reversal-invariant $p+d$ state
near the interface.
It is highly nontrivial how the interaction between these two states
affects the stability of the topological superconductivity.
However, it is expected that as long as the thickness of the region, where
the asymmetric SO interaction is dominant, is sufficiently large compared to
the coherence length of Cooper pairs, the topological superconductivity
should be stable in the vicinity of the interface.
Another point required for the realization of the topological superconductivity 
is the fabrication of the
interface where the electronic structure is not so different from the bulk one.
In the (001) surface of Sr$_2$RuO$_4$, however, it is known 
that a structural phase 
transition occurs and the surface state is ferromagnetic\cite{matzdorf}.
To observe the time-reversal symmetric superconductivity,
a very carefully fabricated sample without such a structural phase transition
is needed.

\section*{Acknowledgments}
The authors have benefited from conversation with M. Sato and S. C. Zhang.
Numerical calculations were partially performed at the 
Yukawa Institute for Theoretical Physics, Kyoto University. 
This work was partly supported by the Grant-in-Aids for
Scientific Research from MEXT of Japan
(Grant No.18540347, Grant No.19014009, Grant No.19014013, Grant No.19052003, 
Grant No.20029013, and Grant No.20102008) 
and the Grant-in-Aid for the Global COE Program 
"The Next Generation of Physics, Spun from Universality and Emergence".
Y. Tada is supported by JSPS Research Fellowships for Young
Scientists.


\end{document}